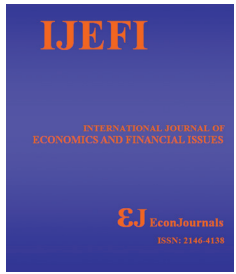

**International Journal of Economics and Financial Issues**

ISSN: 2146-4138

available at http: www.econjournals.com

**International Journal of Economics and Financial Issues, 2016, 6(S7) 109-114.**

Special Issue for "International Soft Science Conference (ISSC 2016), 11-13 April 2016, Universiti Utara Malaysia, Malaysia"

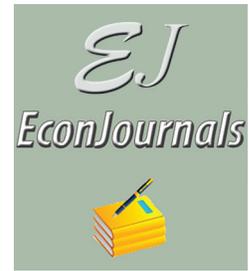

# Determinants of Social-economic Mobility in the Northern Region of Malaysia

**Siti Hadijah Che Mat[1]\*, Mukaramah Harun[2], Nor'Aznin Abu Bakar[3]**

[1]School of Economics, Finance and Banking, College of Business, Universiti Utara Malaysia, 06010 Sintok, Kedah, Malaysia, [2]School of Economics, Finance and Banking, College of Business, Universiti Utara Malaysia, 06010 Sintok, Kedah, Malaysia, [3]School of Economics, Finance and Banking, College of Business, Universiti Utara Malaysia, 06010 Sintok, Kedah, Malaysia.
\*Email: hadijah@uum.edu.my

**ABSTRACT**

Colleting the data through a survey in the Northern region of Malaysia; Kedah, Perlis, Penang and Perak, this study investigates intergenerational social mobility in Malaysia. We measure and analyzed the factors that influence social-economic mobility by using binary choice model ($\log_{it}$ model). Social mobility can be measured in several ways, by income, education, occupation or social class. More often, economic research has focused on some measure of income. Social mobility variable is measured using the difference between educational achievement between a father and son. If there is a change of at least of two educational levels between a father and son, then this study will assign the value one which means that social mobility has occurred. We found that besides father's education, father's attitude and the establishment of a university in the area have also contributed to social mobility of the rural communities.

## 1. INTRODUCTION

In Malaysia, it is crucial to measure the "equality" of society more along the lines of economic social mobility rather than purely through the income or wealth measures. The idea is to have people begin at more or less the same starting point, then proceed in accordance to their own ability and willingness to work. Economic mobility in this sense means that whatever your personal circumstances, you can reach the top of the social and economic ladder. This kind of viewpoint tolerates high income and wealth inequality as the product of a system where everyone has a fair chance.

Intergenerational mobility reflects a host of factors, including inherited traits, investment on human capital (education), social norms and public policies that may influence the individual willingness and ability to seize economic opportunities. If you start off disadvantaged, the odds are stacked against you. For the vast majority of the population, educational attainment and lifetime earnings are as much a product of your background as they are your innate abilities. In other words, where you start from actually matters and intergenerational transfer of wealth skews things considerably. Worse, even a completely equal and meritocratic society will degenerate into a fully unequal society unless there is social or public intervention.

Usually, studies assessing the association between parent's and offspring's incomes focus on pairs of fathers and sons. Ideally, income should be measured by a household's disposable income as this most directly influences the standard of living of individuals (e.g., Chadwick and Solon, 2002; Solon, 2004). In practice, accurate measurement of a household's disposable income is difficult as it should take into account the structure of the household, the extent of female participation in the labor market, as well as the different sources of income (e.g., earnings, assets, welfare). Therefore, most existing studies use some measure of wages. The extent to which the offspring's wage levels reflect those of their parents (the so-called "intergenerational wage elasticity") is taken as a measure of wage persistence across generations or lack of intergenerational wage mobility. The higher this elasticity, the lower is intergenerational wage mobility.

There are a variety of public and private development projects in the rural areas. Educational institutions such as primary schools, secondary schools, colleges, and universities sprouted conspicuously as well as infrastructure facilities such as roads, airports enlarged and upgraded as an international airport and so on. These things are the new engine of growth in the rural communities.

However, what is the position of the rural communities today compared to 10 or 15 years ago? Have their families escaped the cycle of poverty and backwardness?

As we all know, education is an important factor to bring change. Education is one of the highest dimensions of the Malays to experience mobility since colonial times until today. Educational achievement can determine whether a person will follow in their father's footstep as farmers and laborers or secures high positions in public administration or private sector. Success factor in education is associated with the ability to mobilize from lower socio- economic strata to higher strata.

Nevertheless, according to some researchers, the culture of putting an emphasis on education among the rural communities is still low. One very important thing to consider is the awareness of parents to inspire and motivate their children to work hard is still low when compared with other community members who live in the cities.

Of all the studies on poverty, no researchers have ever investigated about social mobility in the rural areas in particular and in-depth. Efforts to understand and explain social mobility problems should include the various dimensions of development as well as the changes that need to be done. Hence, this study decides to examine about what are the factors that contribute or otherwise hinder social mobility from occurring in the rural communities? Are the internal factors at the father's and the family's stage still not strong enough to encourage their children to change? The question of the extent of social mobility occurring in Malaysia is examined based on two indicators, which are the level of education and occupational category between two generations, parents who live in the rural areas and their children. Transformation in social mobility seen based on the level of education that is by examining the highest level of education of two generations, the respondents and their son.

## 2. LITERATURE REVIEW

Based on the definition in the Cultural Dictionary, "social mobility is the ability of individuals or groups to move upward or downward in status based on wealth, occupation, education, or some other social variable." Meanwhile, according to Wikipedia, "social mobility is the movement of individuals or groups of people in social position. It may refer to classes, ethnic groups, or entire nations, and may measure health status, literacy, or education. More commonly, it refers to individuals or families, and their change in income. Social mobility can be the change in status between someone (or a group) and their parents/previous family generations ("inter-generational"); or over the change during one's lifetime ("intra-generational"). It can be "absolute," i.e., total amount of movement of people between classes, usually over one generation (such as when education and economic development raises the socio-economic level of a population); or "relative," which is an estimation of the chance of upward or downward social mobility of a member of one social class in comparison with a member from another class. A higher level of intergenerational mobility is often considered a sign of greater fairness, or equality of opportunity, in a society.

Fields (2000) noted that economic mobility studies are concerned with quantifying the movements of given recipient units through the distribution of economic well-being over time, establishing how dependent one's current economic position is on one's past position. He also mentioned that despite some basic agreement about the concept of economic mobility, there are also some fundamental disagreements because the term income mobility conjures up very different ideas in people's minds. He stated firmly that the mobility literature is plagued by people talking past one another because one person's idea of mobility is not another's.

According to Benabou and Ok (2001), equality of opportunity provides a very natural approach to the evaluation of mobility processes, so natural that there is in fact no need for special concepts or indices to measure it. One cares about mobility not because income movements are intrinsically valuable, but primarily because of the view -or the hope- that it helps attenuate the effects of disparities in initial endowments or social origins, on future income prospects. From this view of mobility as an equalizer of opportunities (but not necessarily of outcomes), it follows quite naturally that one should measure it precisely by the extent to which it achieves such leveling.

The relationship between income inequality and intergenerational income persistence is not straightforward as various channels are at work, sometimes in opposite directions. On the one hand, countries with a wide distribution of income are also likely to be those where the returns to education are relatively high. Inequality could increase intergenerational mobility by enhancing incentives to undertake effort, e.g., by working longer hours or by strengthening incentives to undertake education, which could result in more investment in education if financial markets are sufficiently developed.

With returns to education likely to be higher in more unequal societies and with incentives to acquire additional education stronger in countries where the "pay-off" from doing so is relatively larger, social mobility could be higher in countries where income or wage dispersion is higher. Educational opportunities for the whole family can transmit the motivation to succeed to children. Lillard and Kilburn (1995) showed that education regimes where access to education is unfavorable to lower income families adversely affect intergenerational mobility. Solon (1999) theoretical model reveals that a more progressive public investment in human capital tends to increase mobility. Another theoretical model by Davies et al. (2004) affirms that "starting from the same inequality, mobility is higher under public than under private education." However, an empirical study of Britain by Blanden et al. (2005) found that "the big expansion in university participation has tended to benefit children from affluent families more and thus reinforce immobility across generations."

Louw et al. (2006) investigated the role that parents' education plays in children's human capital accumulation. The study analyses patterns of educational attainment in South Africa during the period 1970-2001, asking whether intergenerational social mobility has improved. It tackles the issue in two-way, combining extensive descriptive analysis of progress in educational attainment with a more formal evaluation of intergenerational social mobility using indices constructed by Dahan and Gaviria (1999) and Behrman (1998). Both types of analysis indicate that intergenerational social mobility within race groups improved over the period, with the indices suggesting that South African children are currently better able to take advantage of educational opportunities than the bulk of their peers in comparable countries. However, significant racial barriers remain in the quest to equalize educational opportunities across the board for South African children.

Causa et al. (2009) examines the potential role of public policies and labor and product market institutions in explaining observed differences in intergenerational wage mobility across 14 European OECD countries. Their empirical results show that education is one important driver of intergenerational wage persistence across European countries. There is a positive cross-country correlation between intergenerational wage mobility and redistributive policies, as well as a positive correlation between wage-setting institutions that compress the wage distribution and mobility.

Azevedo and Bouillon (2010) stated that while intergenerational education mobility have improved in recent decades, which may increase income mobility for younger cohorts, overall, the Latin American region still presents lower intergenerational social mobility. Previous studies suggest that these results might be associated to social exclusion, low access to higher education, public policies and labor market discrimination. Kenway et al. (2005) found that class origins were the key to children's occupational outcome but that having economic assets in the home, and having a highly qualified mother were also very important.

According to d'Addio (2007) parental background can influence their offspring's wages in various ways. In very general terms, parental background can affect these wages by boosting both the offspring's labor productivity and their successful insertion in the labor market. One way in which children's productivity, and hence their future incomes, can be enhanced is through the ability of parents to invest in their offspring's human capital. However, wealth and assets passed on from one generation to another, the inheritance of traits that are important for economic success, such as propensities to undertake education, work ethics and risk-related behaviors, as well as local conditions such as growing up in advantaged neighborhoods are other important factors explaining the transmission of income across generations.

Eberharter (2013) used data from the German Socio-Economic Panel, the Panel Study of Income Dynamics, and the British Household Panel Survey to analyze the hypotheses that the extent and the determinants of intergenerational income mobility and the relative risk of poverty differ with respect to the existing welfare state regime, family role patterns, and social policy design. The empirical results indicate a higher intergenerational income elasticity in the United States than in Germany and Great Britain, country differences concerning the influence of individual and parental socio-economic characteristics, and social exclusion attributes on intergenerational income mobility and the relative risk of poverty.

Causa and Johansson (2010) noted that public policies such as education and early childcare play a role in explaining observed differences in intergenerational social mobility across countries. In addition, their study also found a positive cross-country correlation between intergenerational social mobility and redistributive policies.

## 3. METHODOLOGY

This study involves four states in the north Peninsular Malaysia which are Perlis, Kedah, Penang and Perak. All respondents are located in the rural areas. The sampling frame was obtained from the Statistics Department, Kuala Lumpur. Though the originally given sample was 400, after undergoing data refining process, the total number suitable for analysis was only 333. All these respondents met the study criteria which is a father who is 50-year-old and above and have at the very least one son who is working.

In estimating the determinants that influence social-economic mobility, we employ a binary choice model based on the maximum likelihood method. Dummy dependence variable (of 0 and 1) is used. The value one is given if there is social mobility occurring between a father and his son. The social mobility variable is measured using the difference between the educational achievement between a father and his son. If there is a change of at least of two educational levels between a father and his son, then this study will assign the value one which means that social mobility has occurred. For instance, if the father has a primary school education and the son has at least a higher secondary education (SPM), then it is declared that social mobility has occurred.

Meanwhile, the value zero shows there has been no change in the educational level between a respondent and his son. For example, if a father is a SPM holder and his son is also a SPM holder, then it is said that social mobility does not exist. Nevertheless, if the father has a tertiary level education and the son also has the same level of education, then this study will assign the value one, meaning social mobility exists.

The $\log_{it}$ model used in this study is specified as follows:

Latent variable specification:

$$Y_i^* = \beta X_i + u_i \tag{1}$$

Where,
$Y_i = 1$ (mobility) if $Y_i^* > 0$,

$Y_i = 0$ (no-mobility) if $Y_i^* \leq 0$,
$u_i$ = Error term,
$\beta$ = Estimated parameter,
$X_i$ = Vector of independent variables.

The error term, $u_i$, is assumed to be logistically distributed. The probability of inter-generation $i$ being mobile or otherwise, is postulated to depend on nine independent variables which are Father education level (*Edu_Father*); father attitude (*Att_Father*); father community involvement (*Inv_Community*); asset ownership in the family (*Asset*); existence of a university in vicinity of the respondent's house (*Avaiable_Uni*); Distance of respondent's house to town centre (*Near_Town*); Distance of respondent's house to highway (*Near_Highway*); Distance of respondent's house to bus station (*Near Bus Station*) and Location of respondent's house to tourism centre (*Near_TourismLoc*).

$$Pr(Y_i = 1|x_i) = F(x_i'\beta) = \frac{\exp(x_i'\beta)}{1+\exp(x_i'\beta)}, \quad (2)$$

Where,
$x_i' = [Edu\_Father_i, Att\_Father_i, Inv\_Community_i, Asset_i, Avaiable\_Uni_i, Near\_Town_i, Near\_highway_i, Near\ Bus\ Station_i, Near\_Tourism\ Loc_i]$.

where;
$X_1$: Father education level (Edu_Father)
$X_2$: Father attitude (Att_Father)
$X_3$: Father community involvement (Inv_Community)
X3: Asset ownership in the family (Asset)
$X_4$: Existence of a university in vicinity of the respondent's house (Avaiable_Uni)
$X_5$: Distance of respondent's house to town centre (Near_Town)
$X_6$: Distance of respondent's house to highway (Near_Highway)
$X_7$: Distance of respondent's house to bus station (Near Bus Station)
$X_8$: Location of respondent's house to tourism centre (Near_TourismLoc).

Equation (2) is used to estimate the probability of occurrence of social mobility. It is worth noting that the sign of the estimated parameter is already sufficient to conclude whether the independent variable has a positive or negative impact on the dependent variable (Wooldridge, 2002). In addition, the impact of the independent variables on the dependent variable could be examined by looking at the odds ratio. Given the value of the independent variables, the estimated value for the dependent variable could be interpreted as the probability of the respondent gaining mobility (Greene, 2000; Maddala, 1983).

## 4. FINDING

This field study data was obtained starting from early 2015. The rural communities selected as respondents are the head of the family, that is fathers aged 50 years and older and having a son who works aged 25 years or older. A total of 333 people from the rural communities were selected based on the selection made by the Department of Statistics Malaysia. Perak has the highest number of respondents with 151 respondents (45%). This is followed by the states of Kedah with 138 respondents (41%, Perlis and Pulau Pinang which recorded 7% of respondents each.

**Table 1: Level of education of father and son**

| Level of education | Frequency (%) | |
|---|---|---|
| | Father | Son |
| No education | 107 (32.1) | 1 (0.3) |
| Primary school | 91 (27.3) | 6 (1.8) |
| Secondary school | 130 (39.0) | 198 (59.5) |
| Tertiary (College and University) | 5 (1.5) | 128 (38.4) |
| Total | 333 (100) | 333 (100) |

Table 1 shows information about the formal education possessed by the respondents who are fathers and the respondents' son. The level of formal education is divided into four classifications, which are no schooling, primary school, lower secondary school (SRP/PMR), higher secondary school (SPM), tertiary education (diploma/degree).

This study examines the highest level of formal education attained between two generations, that is, the education level of parents and children. As shown in Table 1, the level of educational attainment by two generations, that is between parents (the new era of independence) and child (around 25-30 years earlier) shows a significant improvement. In accordance with the life in the new era of independence with rampant deprivation and limited access to education, 32.1% of the fathers have never received any formal education. Moreover, only 67.9% of the fathers have gone through a formal education system where 39% have attained secondary education level and 1.5% has tertiary education.

There is a noticeable increase in the level of education obtained by the sons where almost 100% of them have received formal education. Moreover, only 0.3% have never received any formal education while 1.8% have received primary education. Almost 98% of them have secondary level of education and above. In fact, almost 40% of the son has received a tertiary education.

Generally, the study has found that there has been a transformation in terms of mobility in rural communities based on the educational aspect that is achieved by the two generations under study, the generations of parents and children. Mobility by level of education is recognized in importance as a key prerequisite for achieving a better life for the rural communities. This reflects that people have become more aware of the importance of formal education in life. In addition, through the well-organized national education system, rural people are able to obtain formal education.

Table 2 presents the details of the factors that have influenced the position and the changing patterns of social mobility of the rural population. The analysis was based on five factors: The father's level of education, the father's attitude, the father's involvement in the community, ownership of assets, and the provision of space and opportunity by the government.

Mat, et al.: Determinants of Social-economic Mobility in the Northern Region of MalaysiaTable 2: Determinants of social mobility in Malaysia: Log$_{it}$ model analysis

| Dependent variable | | | |
|---|---|---|---|
| Son's education (Social mobility) | Son's education at high level (Binary) (Yes=1, No=0) | | |
| Independent variable | Estimated coefficient | | |
| | Parameter | Standard error | Odds ratio |
| Constant | −4.7311 | 1.3088 | |
| Edu_Father | 0.6693*** | 0.2308 | 0.9530 |
| Att_Father | 2.1264** | 1.0097 | 8.3846 |
| Inv_Community | −0.5335 | 1.534 | 0.5865 |
| Asset | 0.0059 | 0.0417 | 1.0059 |
| Avaiable_Uni | 0.8015*** | 0.2333 | 2.2290 |
| Near_Town | −0.0381 | 0.2722 | 0.9626 |
| Near_Highway | −0.2281 | 0.2763 | 0.7961 |
| Near_Bus station | −0.1304 | 0.1279 | 0.8778 |
| Near_Tourism | 0.1464 | 0.2055 | 1.1576 |
| Log likelihood | | −205.1975 | |
| Number of observation | | 331 | |
| LR Chi-square (9) | | 30.39 | |
| P>Chi-square | | 0.0004 | |
| Pseudo R$^2$ | | 0.0689 | |

**Significant at 5% level, ***Significant at 1% level

In this analysis, the social mobility variable is measured by using the difference of the level of educational attainment between the respondent and his son. In using the log$_{it}$ model analysis, a dichotomous dependent variable or a variable that has only two decision values has been set, whether one, that is the existence of social mobility and zero there is no social mobility. As described in the methodology section of the study, in the event of at least they are two changes in the level of education between the respondents and his son, then this study will give the value one and vice versa.

The log$_{it}$ regression analysis uses nine independent variables which are education level of father (*Edu_Father*); attitude of the father (*Att_Father*); father's involvement in the community (*Inv_Community*); ownership of assets in the family (*Asset*); existence of a university in the immediate vicinity of the respondent's house (*Available_Uni*); distance of respondent's house to the town centre (*Near_Town*); distance of respondent's house to the highway (*Near_highway*); distance of respondent's house to the bus stop (*Near Bus Station*) and position of the respondent's house with tourism centre (*Near_Tourism_Loc*).

This study uses the value of the odds ratio to illustrate the likelihood of a change in social mobility. The odds ratio value is a figure that reflects the value of the choice of whether there is a change in the level of social mobility or not.

The factor of father's education level (*Edu_Father*) shows significant value at the 1% significance level in determining the existence of social mobility. The positive correlation means that the higher the father's education level, there is a higher probability for the occurrence of social mobility. This situation is the same as what was anticipated at the beginning of the study. This situation may arise because when a father who is also a leader in the family has high education, then he (the father) would feel more responsible for ensuring that his children also acquire an education level that is equally high or higher than himself. The estimation results also show that the effect of a change in the father's level of education (*Edu_Father*) as indicated by the odds ratio in the father's education level is 1.9530. This means that one educational level increase in the father's education will cause the odds value of the occurrence of social mobility to increase by a factor of 1.95, ceteris paribus.

The factor of father's self-identity or attitude (*Att_Father*) is also significant at the 5% level of significance in determining the occurrence of social mobility. There are four elements of identity measured using a Likert scale considered in the analysis which are the motivation to work hard, willingness to learn new things, willingness to take risks and not easily demotivated.

The study found that there is a positive relationship between these factors with social mobility and it is in line with the expectations at the beginning of the study. This means that the higher the value or amount of a father's self-identity, the higher the probability of the occurrence of a change in the level of the son's education and vice versa.

This situation may arise because when a father who is also a leader in the family has a positive self-esteem, then this father's attitude will lead to the formation of positive behavior of his children as well. The study found that the effect of changes in the father's self-esteem father (*Att_Father*) shown by the odds ratio of self-identity is 8.3846. This means that a one per cent increase in a father's self-esteem will cause the odds of the occurrence of social mobility to be increased by a factor of 8.38, ceteris paribus.

The establishment of a college or university that is close to the respondent's residence is found to have a statistically significant relationship at the 1% level of significance, to the probability of the existence of social mobility. The establishment of an institution of higher learning indirectly affect parents' awareness of the importance of education. Activities or education and community service programmes carried out by universities and their students in the local community provide exposure to parents and their children about higher education. For example, universities often carry out motivational programmes to SPM and STPM students to foster their interest to continue learning and succeeding in their studies.

## 5. CONCLUSION

The study was based on a theoretical stance that no one single factor can explain social mobility. In contrast, changes in the mobility form should be seen within the framework of a multi-causal or multi-factoral analysis. For social mobility to occur, the person needs a combination of the driving factors, in particular, the factors of education, occupation, attitude, asset ownership and the role of government simultaneously. In examining the factors that influence socio-economic mobility, we used log$_{it}$ model with nine independent variables. The odds ratio value is used to analyze the probability of changes that occurs in the factor that leads to social mobility. Out of the nine variables, only three are significant

International Journal of Economics and Financial Issues | Vol 6 • Special Issue (S7) • 2016     113

(father's education, father's attitude, the existence of a university in the near vicinity).

Those that are in the high mobility position are fathers who have strong spirits and internal ability compared with fathers who experienced decreased mobility. Strong internal ability is seen through high self-regard to change. The importance of development projects implemented by the government is perceived as one of the main factors that influence social mobility where 99.28% of the respondents agree that the existence of higher learning institutes helps to stimulate social mobility.